\documentclass[twocolumn,aps]{revtex4}
\usepackage{graphicx}
\newcommand{\beq}{\begin{equation}}
\newcommand{\eeq}{\end{equation}}
\newcommand{\beqa}{\begin{eqnarray}}
\newcommand{\eeqa}{\end{eqnarray}}
\begin{document}
\title{Self-gravitational solitary waves in astrophysical compact objects}
 \author{A. A. Mamun} \affiliation{Department of Physics,
Jahangirnagar University, Savar, Dhaka-1342, Bangladesh}
\begin{abstract}
The condition for the existence of self-gravitational solitary
waves (SGSWs), and their polarity in any astrophysical compact
object are theoretically found for the first time. The
pseudo-potential approach, which is valid for arbitrary amplitude
SGSWs, is emplyed. The general analytical results are applied in
white dwarfs and neutron stars to identify the basic features
(polarity, amplitude, and width) of the SGSWs formed in them. It
found for the first time that the SGSWs exist with negative
self-gravitational potential in perturbed states of white dwarfs
and neutron stars. It is also estimated that for their typical
degenerate plasma parmeters, the amplitude and the width of the
SGSWs (moving with the speed 2 cm/s) in white dwarfs are $\sim
-1.5$ ergs/gm and $\sim 50$ cm, respectively, and those of the
SGSWs (moving with the speed $\sim 27$ m/s) in neutron stars are
$\sim -7.76\times 10^4$ Joules/kg, and $5.5$ km, respectively.
\end{abstract}
\maketitle
The astrophysical compact objects (ACOs) are completely different
from other terrestrial bodies because of their (ACOs') extremely
low temperate and extremely high density
\cite{Chandrasekhar1931,Shapiro2004,Horn1991}. They are, in fact,
degenerate quantum plasma
\cite{Manfredi2005,Shukla2006,Markcloud2007,Shukla2011a,Haas2011,Shukla2011b,MAS2016,Brodin2017,MAS2017}
systems containing degeneare light particle species (viz.
electron species with/without positron or/and quark (non-zero
mass \cite{Fraga2005}) species depending on the class of ACOs
under consideration), degnerate heavy particle (compared to
electron mass) species
\cite{Chandrasekhar1931,Koester1990,Koester2002,Shapiro2004}
(viz. $~^{4}_{2}$He or/and $~^{12}_{~6}$C or/and $~^{16}_{~8}$O
species with/without proton or/and neutron species depending on
the class of ACOs under consideration), and non-degenerate species
of low densed heavy elements
\cite{Shapiro2004,Witze2014,Vanderburg2015} (viz. $~^{56}_{26}$Fe
or/and $~^{85}_{37}$Rd or/and $~^{96}_{42}$Mo depending again on
the class of ACOs under consideration).

The degeneracy of the light and heavy particle species (viz.
electron or proton or neutron or $~^{4}_{2}$He or $~^{12}_{~6}$C
or $~^{16}_{~8}$O species) arises due to Heisenberg's uncertainty
principle, $\Delta p\Delta x\ge \hbar/2$  (where $\hbar$ is the
reduced Planck constant, $\Delta p$ is the uncertainty in
momentum of the species, and $\Delta x$ is the uncertainty in
position of the species). This indicates that the momenta of
highly compressed particles are extremely uncertain, since the
particles are located in a very confined space. Therefore, even
though the degenerate plasma is very cold, the plasma particles
must move very fast on average, and give rise to a very high
pressure (known as degenerate pressure), which does not depend on
thermal temperature, but on degenerate particle number density.
This leads us to a conclusion that in order to compress
degenerate particles to an object into a very small space, a
tremendous force (which is the self-gravitational force in ACOs
like white dwarfs, neutron stars, black holes, etc.
\cite{Shapiro2004}) is required to control its particles'
momenta. Thus, if any ACO is disturbed/perturbed due any reason
(viz. merging \cite{Abbott2016} of two small ACOs, fragmentation
\cite{Shapiro2004} of a large ACO, gravitational interaction
\cite{Shapiro2004} among neighboring ACOs,  etc.), the
self-gravitational perturbation mode [in which the compression
(rarefaction) is provided by the self-gravitational attraction
pressure (degenerate pressure)] is developed.  The propagation of
this self-gravitational perturbation (SGPM) cannot be usually
explained by linear analysis. Therefore, for the first time a
theoretical investigation is made on the nonlinear propagation of
this SGPM by employing the pseudo-potential approach, which is
valid for arbitrary amplitude SGPM. It is first time shown here
that the SGPM nonlinearly propagate as self-gravitational
solitonic potential signals (SGSWs) with negative potential in
the ACOs like white dwarfs and neutron stars.

We consider a general degenerate quantum plasma system like an
ACO containing arbitrary number of degenerate species $s$ of
non-inertial light particles
\cite{Chandrasekhar1931,Shapiro2004,Horn1991} (viz. electrons
or/and positrons or/and non-zero mass quarks \cite{Fraga2005},
etc.), degenerate species $j$ of inertial heavy (compared to
electron mass) particles
\cite{Shapiro2004,Horn1991,Koester1990,Koester2002} (viz.
$~^{4}_{2}$He or $~^{12}_{~6}$C or $~^{12}_{~6}$O or/and protons
or/and neutrons, etc.), and non-degenerate species $h$ of
inertial heavy elements \cite{Horn1991,Vanderburg2015,Witze2014}
(viz. $~^{56}_{26}$Fe
 or/and $~^{85}_{37}$Rd or/and
$~^{96}_{42}$Mo, etc.). The heavy elements, viz. $~^{56}_{26}$Fe,
$~^{85}_{37}$Rd, $~^{96}_{42}$Mo, etc., are assumed to be
non-degenerate just because of their low density and high mass.

The nonlinear dynamics of the SGPM, whose phase speed is much
smaller than $C_{e}$ (where $C_{e}=\sqrt{\pi}\hbar
n_{e0}^{1/3}/m_e$ in which $m_e$ is the electron mass, and
$n_{e0}$ is the equilibrium degenerate electron number density)
is described by
\begin{eqnarray}
&&\frac{\partial \psi}{\partial
x}=-\frac{3}{2}\alpha_s^\prime\frac{\partial\rho_s^{\frac{2}{3}}}{\partial
x},
\label{be1}\\
&&\frac{\partial \rho_j}{\partial t} +\frac{\partial}{\partial
x}(\rho_ju_j) = 0,
\label{be2}\\
&&\frac{\partial u_j}{\partial t} +u_j\frac{\partial
u_j}{\partial x}=-\frac{\partial \psi}{\partial
x}-\frac{3}{2}\beta_j\frac{\partial \rho_j^{\frac{2}{3}}}{\partial
x},
\label{be3}\\
&&\frac{{\partial}^2 \psi}{\partial x^2}=\sum_s
\sigma_s(\rho_s-1)+\sum_j \mu_j(\rho_j-1),
 \label{be4}
\end{eqnarray}
where $\rho_s$ ($\rho_j$) is the number density of the degenerate
light (heavy) particle species $s$ ($j$), and is normalized by
$\rho_{s0}$ ($\rho_{j0}$) in which $\rho_{s0}$ ($\rho_{j0}$) is
the equilibrium mass density of the degenerate particle species
$s$ ($j$);  $u_j$ is the degenerate fluid speed of the species
$j$, and is normalized by $C_q$ in which $C_q=(\sqrt{\pi}\hbar
n_{e0}^{1/3}/m_q)$, and $q$ represents the heaviest particle
species among all the degenerate particle species (e. g. $m_q$ is
the mass of $~^{12}_{~6}$C for white dwarfs and of $~^{4}_{2}$He
for neutron stars); $\psi$ is the self-gravitational potential,
and is normalized by $C_q^2$;
$\alpha_s^\prime=(m_q/m_s)^2(n_{s0}/n_{e0})^{2/3}$, $\beta_j=
(m_q/m_j)^2(n_{j0}/n_{e0})^{2/3}$, in which $m_s$ ($m_j$) is the
mass of the degenerate particle species $s$ ($j$), and $n_{s0}$
($n_{j0}$) is the equilibrium number density of the degenerate
particle species $s$ ($j$); $t$ is the time variable normalized by
$\tau_q=(4\pi G\rho_{q0})^{-1/2}$; $x$ is the space variable
normalized by $L_q=C_q\tau_q$; $\sigma_s=\rho_{s0}/\rho_{q0}$,
and $\mu_j=\rho_{j0}/\rho_{q0}$ in which $\rho_{q0}$ is the mass
density of the degenerate species $q$ (e. g. $~^{12}_{~6}$C for
white dwarfs and $~^{4}_{2}$He for neutron stars). We note that
$j=h$ when $\beta_j=0$. The  typical approximate values of the
physical quantities, viz. $m_q$, $m_h$, $\rho_{e0}$ $\rho_{q0}$,
$\rho_{n0}$, $C_q$, $\omega_{Jq}$, and $L_q$, corresponding to
white dwarfs and neutron stars are tabulated in Table I.
\begin{table}[!h]
\caption{Typical approximate values of the physical quantities,
$m_q$, $m_h$, $\rho_{e0}$, $\rho_{q0}$, $\rho_{n0}$, $C_q$, $L_q$,
and $\omega_{Jq}$ corresponding to white dwarfs
\cite{Shapiro2004,Horn1991,Koester2002,Vanderburg2015} and neutron
stars \cite{Shapiro2004,Horn1991,Witze2014}.}
\begin{tabular}{|c|c|c|c|}
\hline
Physical quantity                     & White dwarfs & Neutron stars \\
\hline $m_q$ (gm)                     & ${\rm 12m_p\footnote{
where ${\rm m_p (=1.67\times10^{-24}~gm)}$ is the proton mass.}}$      & ${\rm 4m_p}$ \\
\hline
$m_h$ (gm)                            & ${\rm 56m_p}$                  & ${\rm 85m_p}$\\
\hline
$\rho_{e0}~{\rm (gm~cm^{-3}})$        & ${\rm 1.82\times 10^3}$        & ${\rm 9.10\times 10^{11}}$\\
\hline
$\rho_{p0}~{\rm (gm~cm^{-3}})$        &  $0$                           & ${\rm 1.67\times 10^{13}}$\\
\hline
$\rho_{n0}~{\rm (gm~cm^{-3}})$        &  $0$                           & ${\rm 1.53\times 10^{14}}$\\
\hline
$\rho_{g0}~{\rm (gm~cm^{-3}})$        & ${\rm 6.35\times 10^6}$        & ${\rm 3.17\times 10^{15}}$\\
\hline
$\rho_{h0}~{\rm (gm~cm^{-3}})$        & ${\rm 3.60 \times 10^5}$       & ${\rm 2.50\times 10^{13}}$\\
\hline
$C_q~{\rm (cm~s^{-1}})$               &  ${\rm 1.17\times 10^6}$       & ${\rm 2.78\times 10^9}$\\
\hline
$\omega_{Jq}~{\rm (s^{-1}})$          &  ${\rm 2.31\times 10^{-2}}$    & ${\rm 5.15\times 10^2}$\\
\hline
$L_q~{\rm (cm})$                      &  ${\rm 5.07\times 10^7}$       & ${\rm 5.40\times 10^6}$\\
\hline
\end{tabular}
\end{table}
We note that $\rho_{h0}$ in both white dwarfs and neutron stars
are determined by the charge neutrality condition, and that
$m_n\simeq m_p$.

We now study the possibility for the formation of the SGSSs, and
their basic features by the pseudo-potential approach
\cite{Bernstein1957,Cairns1995}, which is valid for arbitrary
amplitude SGSWs. To do so, we first assume that all dependent
variables in (\ref{be1})-(\ref{be4}) depend only on a single
variable $\xi=x-{\cal M}t$, where ${\cal M}$ is the speed of the
the SGSWs, and is normalized by $C_q$, and $\xi$ is normalized by
$L_q$. On using (\ref{be1})-(\ref{be4}), this assumption leads us
to express $\rho_s$ and $\rho_j$ (under the steady state
condition) as
\begin{eqnarray}
&&\rho_s=\left(1-\frac{2}{3}\alpha_s\psi\right)^{\frac{3}{2}},
\label{be5}\\
&&\rho_j=\mu_j\left[1 - b_j + b_j\sqrt{1 - \frac{2\psi}{{\cal
M}^2b_j^2}}~\right]^{-1},
\label{be6}\\
&&\frac{{\partial}^2 \psi}{\partial \xi^2}=\sum_s
\sigma_s(\rho_s-1)+\sum_j \mu_j(\rho_j-1),
\label{be7}
\end{eqnarray}
where $\alpha_s=1/\alpha_s^\prime$, and $b_j =(1-\beta_j/{\cal
M}^2)/(1+5\beta_j/3)$. We note that the species $j$ includes the
species $q$ and $h$, and that $\rho_j=\rho_q$ when $j$ represents
the heaviest degenerate species (e. g. $~^{12}_{~6}$C for white
dwarfs \cite{Koester1990,Koester2002}, and $~^{4}_{2}$He for
neutron stars \cite{Shapiro2004,Horn1991}), and $\rho_j=\rho_h$
when $j$ represent non-degenerate species ($\beta_j=0$) of heavy
elements (e. g. $~^{56}_{26}$Fe in white dwarfs
\cite{Horn1991,Vanderburg2015} and $~^{85}_{37}$Rd in neutron
stars \cite{Horn1991,Witze2014}).

We first substitute (\ref{be5})  and  (\ref{be6}) into
(\ref{be7}),  and then integrate the resulting equation over
$\xi$ to obtain an energy integral in the form
\cite{Bernstein1957,Cairns1995}
\begin{eqnarray}
\frac{1}{2}\biggr(\frac{d\psi}{d\xi}\biggr)^{2}+V(\psi)=0.
\label{EI}
\end{eqnarray}
This energy integral represents the motion of a pseudo particle
of unit mass moving with pseudo-speed $d\psi/d\xi$ in a
pseudo-potential $V(\psi)$, where $\psi$  is the pseudo-position
and  $\xi$ is the pseudo-time \cite{Bernstein1957}. The
pseudo-potential $V(\psi)$ appeared in (\ref{EI}) is given by
\begin{eqnarray}
&&\hspace*{-4mm}V(\psi)=-\sum_s\delta_s\left[\frac{3}{5\alpha_s}-\psi-\frac{3}{5\alpha_s}\left(1
-\frac{2}{3}\alpha_s\psi\right)^{\frac{5}{2}}\right]\nonumber\\
&&\hspace*{-4mm}-\sum_j\mu_j\left[b_j{\cal M}^2-\psi-b_j{\cal
M}^2Y_j+(1-b_j){\cal M}^2Z_j\right],
\label{PP}
\end{eqnarray}
where $b_j=(1-\beta_j/{\cal M}^2)/(1+5\beta_j/3)$, $Y_j=[1 -
2\psi/(b_j^2 {\cal M}^2)]^{1/2}$, $Z_j=\ln(1-b_j+b_jY_j)$, and the
integration constant is chosen in such a way that $V(\psi)=0$ at
$\psi=0$.  We note that the energy integral defined by (\ref{EI})
with the pseudo-potential $V(\psi)$ defined by (\ref{PP}) is valid
for any ACO containing arbitrary number of degenerate species $s$
of light particles (viz. electrons, positrons, etc.), arbitrary
number of degenerate species $j$ of heavy particles (viz.
protons, neutrons, $~^{4}_{2}$He nuclei, $~^{12}_{~6}$C nuclei,
etc.), and arbitrary number of non-degenerate species $h$
($\beta_h=0$) of heavy elements (viz. $~^{56}_{26}$Fe,
$~^{85}_{37}$Rd, $~^{96}_{42}$Mo, etc.). The Taylor series
expansion of $V(\psi)$ around $\psi=0$ is
\begin{eqnarray}
V(\psi)=\frac{1}{2}C_2\psi^2+\frac{1}{2}C_3\psi^3+\cdot \cdot
\cdot,
\label{TE}
\end{eqnarray}
where
\begin{eqnarray}
&&\hspace*{-6mm}C_2=\sum_s\delta_s\alpha_s-\sum_j\frac{\mu_j\left(1+\frac{5}{3}\beta_j\right)}
{({\cal M}^2-\beta_j)},
\label{C2}\\
&&\hspace*{-6mm}C_3=-\frac{1}{9}\sum_s\delta_s\alpha_s^2\nonumber\\
&&-\frac{1}{3}\sum_j\frac{\mu_j\left(1+\frac{5}{3}\beta_j\right)^2[{\cal
M}^2\left(3+\frac{5}{3}\beta_j\right)-2\beta_j]}{({\cal
M}^2-\beta_j)^3}.
\label{C3}
\end{eqnarray}
It is obvious from (\ref{TE}) that $V(0)=0$ and
$(dV/d\psi)|_{\psi=0}=0$.  This means  that (i) the energy
integral defined by (\ref{EI}) with the pseudo-potential
$V(\psi)$ defined by (\ref{PP}) gives rise to the existence of
SGSWs if and only if ($d^2V/d\psi^2)|_{\psi=0}<0$, i. e. $C_2<0$,
and thus $C_2|_{{\cal M}={\cal M}_c}=0$ yields ${\cal M}_c$ (the
critical value of ${\cal M}$ bellow which the SGSWs exist); (ii)
the SGSWs exist with $\psi>0$ if $C_3|_{{\cal M}={\cal M}_c}>0$,
and with $\psi<0$ if $C_3|_{{\cal M}={\cal M}_c}<0$.

To find the expression for ${\cal M}_c$ and to examine whether
$C_3|_{{\cal M}={\cal M}_c}>0$ or $C_3|_{{\cal M}={\cal M}_c}<0$,
for an example an example, we first consider white dwarfs, which
can be assumed to consist of degenerate species $s$ of light
particles like \cite{Shapiro2004} electrons, degenerate species
$q$ of heavy (compared to electron mass) particles like
\cite{Koester2002,Shapiro2004} $~^{12}_{~6}$C, and non-degenerate
species $h$ of heavy element like \cite{Vanderburg2015}
$~^{56}_{26}$Fe). Thus, using (\ref{C2}) and (\ref{C3}), the
conditions for the existence of SGSWs in white dwarfs become
\begin{eqnarray}
&&\hspace*{-6mm}{\cal M}^2<{\cal
M}_c^2=\frac{1}{2\alpha_e\delta_e}\left[\beta_1
+\sqrt{\beta_1^2-4\alpha_e\beta_q\delta_e\mu_h}\right],
\label{cond1}\\
&&\hspace*{-6mm}C_3|_{{\cal M}={\cal
M}_c}=C_3^c=-\frac{1}{9}\delta_e\alpha_e^2-\frac{\mu_h}{{\cal M}_c^4}\nonumber\\
&&\hspace*{6mm}-\frac{1}{3}\frac{\left(1+\frac{5}{3}\beta_1\right)^2[{\cal
M}_c^2\left(3+\frac{5}{3}\beta_q\right)-2\beta_l]}{({\cal
M}_c^2-\beta_q)^3},
\label{cond2}
\end{eqnarray}
where $\beta_1=1+\mu_h+\beta_q(\alpha_e\delta_e+ 5/3)$. It is
obvious from (\ref{cond2}) that we cannot analytically predict
$C_3^c<0$ for any possible value of $\beta_q$, which is $0.29$
for $~^{12}_{~6}$C. The numerical variation of $C_3^c$ with
$\beta_q$ (for its range from $0$ to $1$) implies that $C_3^c<0$
for $\beta_q=0-1$. Similarly, we can also use (\ref{C2}) and
(\ref{C3}) to verify that the the conditions (viz. ${\cal
M}<{\cal M}_c$ and $C_3^c<0$) for the existence of the SGSWs in
neutron stars are satisfied. However, the expressions of ${\cal
M}_c$ and $C_3^c$ for neutron stars are more complex than those
for white dwarfs. This is because the neutron stars can be
assumed to consist of a degenerate species of light
particles\cite{Shapiro2004} (viz. electrons), three degenerate
species of heavy (compared to electron mass) particles
\cite{Shapiro2004} (viz. protons, neutrons, light nuclei like
$~^{4}_{2}$He), and a non-degenerate ($\beta_h=0$) species of
heavy element like \cite{Witze2014} $~^{85}_{37}$Rd. We note that
in neutron stars $j=q$ for $~^{4}_{2}$He species, and that
$\beta_j$ is $0.74$, $0.93$, and $0.97$ for $~^{4}_{2}$He,
proton, and neutron species, respectively.

\begin{figure}[!h]
\centering
\includegraphics[width=8cm]{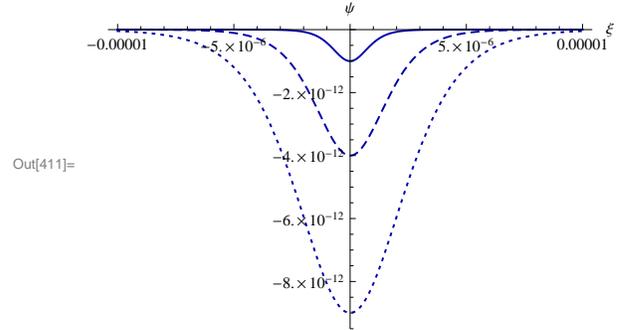}
\caption{Showing the SGSWs formed in white dwarfs for ${\cal
M}{\rm =1.0\times 10^{-6}}$ (solid curve), ${\cal M}{\rm
=2.0\times 10^{-6}}$ (dotted curve),  and ${\cal M}{\rm
=3.0\times 10^{-6}}$ (dash curve).} \label{Fig1}
\end{figure}

\begin{figure}[!h]
\centering
\includegraphics[width=8cm]{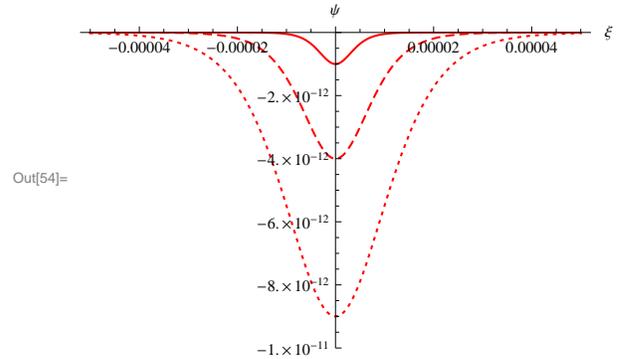}
\caption{(Color online) Showing  the SGSWs formed in neutron stars
for ${\cal M}{\rm =1.0\times 10^{-6}}$ (solid curve), ${\cal
M}{\rm =2\times 10^{-6}}$ (dotted curve), and ${\cal M}{\rm
=3.0\times 10^{-6}}$ (dash curve).} \label{Fig2}
\end{figure}

To examine the basic features (amplitude and width) of  these
SGSWs, we limit $V(\psi)$ up to the terms containing $\psi^3$
(since $\psi\ll 1$ always valid because of its normalization by
$C_q^2$, which is $1.36\times 10^{12}{\rm cm^2~s^{-2}}$ for white
dwarfs, and $7.73\times 10^{18}{\rm cm^2~s^{-2}}$ for neutron
stars). This limit allows us to express (\ref{EI}) as
 \begin{eqnarray}
\frac{d\psi}{d\xi}=\psi\sqrt{-C_2-C_3\psi}.
\label{sa1}
\end{eqnarray}
We now look for the solitonic signal solution of this first order
ordinary nonlinear differential equation (\ref{sa1}), and  under a
localized boundary condition, viz. $\psi\rightarrow 0$ at
$\xi\rightarrow \pm \infty$,  its stationary solitonic signal
solution can be written as
\begin{eqnarray}
\psi=-\left(\frac{C_2}{C_3}\right){\rm Sech^2}(\sqrt{-C_2}\xi).
\label{sa2}
\end{eqnarray}
This represents a solitonic signal solution of (\ref{sa1}), and
means that the SGSWs with $\psi<0$ exist in the degenerate plasma
system under consideration since $C_2<0$ and $C_3<0$. We have
graphically shown the SGSWs that can be formed for the parameters
corresponding to white dwarfs and neutron stars. These are
displayed in figures \ref{Fig1} and \ref{Fig2},
which clearly indicates that the SGSWs exist with $\psi<0$) in
ACOs (like white dwarfs and neutron stars) and that the magnitude
of the amplitude and the width of the SGSWs increase with the
rise of ${\cal M}$ within the range of ${\cal M}_c<{\cal M}<0$.
We have also numerically solved (\ref{EI}) [with the
pseudo-potential $V(\psi)$ defined by (\ref{PP})] for the
parameters given in Table I, and used in figures \ref{Fig1} and
\ref{Fig2}, and have found the existence of the SGSWs, which are
completely identical to those shown in figures \ref{Fig1} and
\ref{Fig2}.

To summarize, we have considered a general self-gravitating
degenerate quantum plasma system [containing an arbitrary number
of degenerate species of non-inertial light particles, (viz.
electrons, positrons, non-zero mass quarks, etc.), arbitrary
number of degenerate species of inertial heavy (compared to
electron mass) particles (viz. protons, neutrons, nuclei of light
elements like $~^{4}_{~2}$He, $~^{12}_{~6}$C, $~^{16}_{~8}$O,
etc.), and an arbitrary number of non-degenerate species of
inertial heavy elements (viz. $~^{56}_{26}$Fe, $~^{85}_{37}$Rd,
$~^{96}_{42}$Mo, etc.)], and have found the conditions for the
existence of the SGSWs in ACOs. The pseudo-potential approach,
which is valid for the arbitrary amplitude SGSWs, is used. The
general analytical results have been applied in white dwarfs
(containing a degenerate electron species \cite{Shapiro2004}, a
degenerate $~^{12}_{~6}$C species \cite{Koester2002,Shapiro2004},
and a non-degenerate $~^{56}_{26}$Fe species
\cite{Vanderburg2015}) and neutron stars (containing degenerate
electron, proton, neutron, and $~^{4}_{2}$He species
\cite{Shapiro2004}, and a non-degenerate $~^{85}_{37}$Rd species
\cite{Witze2014}) to identify the basic features (viz. polarity,
amplitude, width, etc.) of the SGSWs, which are found to be
formed in them. It is shown that the SGSWs exist with negative
self-gravitational potential in both white dwarfs and neutron
stars for ${\cal M}_c<{\cal M}<0$. It is observed that the speed
($V_0$), magnitude of the amplitude ($|\psi_m|$) and the width
($\delta$) of the SGSWs (in both white dwarfs and neutron stars)
increase with the rise of their normalized speed (${\cal M}$)
within the range $0<{\cal M}<{\cal M}_c$. It has been carefully
checked that the analysis for arbitrary amplitude SGSWs gives the
same results as that for small amplitude SGSWs does since
$|\psi|\ll 1$ always valid because of its normalization by
$C_q^2$, which is $1.36\times 10^{12}{\rm cm^2~s^{-2}}$ for white
dwarfs, and $7.73\times 10^{18}{\rm cm^2~s^{-2}}$ for neutron
stars. It is also observed that the dimensional (non-normalized)
values of $V_0$, $\psi_m$, and $\delta$ in neutron stars are much
higher than those in white dwarfs. It has been estimated that for
${\cal M}=10^{-6}$ and for the parameters given in Table I, $V_0$,
$|\psi_m|$, and $\delta$ are $\sim 2$ cm/s, $\sim 1.5$ ergs/gm,
and $\sim 50$ cm, respectively, in white dwarfs, and are $\sim
27$ m/s, $\sim 7.76\times 10^4$ Joules/kg, and $5.5$ km,
respectively, in neutron stars.

Recent discovery \cite{Abbott2016} of the gravitational waves
\cite{Abbott2016,Kurkela2016,Ho2016} produced by merging of two
black holes, and the results of the present investigation
indicate that the signatures of the nonlinear structures like
SGSWs exist in ACOs like white dwarfs and neutron stars. The
SGSWs are associated with the perturbation mode generated  in
ACOs (viz. white dwarfs and neutron stars) due to the departure
from their equilibrium states by any event (viz. merging of two
small ACOs \cite{Abbott2016,Kurkela2016,Ho2016}, splitting up a
large ACO according to the Chandrasekhar limit
\cite{Chandrasekhar1931,Shapiro2004}, gravitational interaction
among nearby ACOs \cite{Shapiro2004}, etc.) or by any other
reasons.

The degenerate quantum plasma system considered here is
generalized to an arbitrary number of species of degenerate light
non-inertial particles, degenerate heavy (compared to electron
mass) inertial particles, and non-degenerate inertial heavy
nuclear species of heavy elements. The degenerate quantum plasma
system is also very general from the view of arbitrary mass
densities of all the species comprising the system. Therefore, the
investigation presented here can be applied for any ACO, and can
be very useful in understanding the basic features of nonlinear
localized structures (viz. SGSWs) in any ACO (viz. white dwarfs,
neutron stars, and black holes \cite{Shapiro2004}).

\end{document}